\documentclass[a4paper]{jpconf}
\usepackage{graphicx}
\usepackage{url}
\begin{document}
\title{Extracting the impact parameter dependence of the nPDFs from the EKS98 and EPS09 global fits}

\author{I Helenius$^{1,2}$, K J Eskola$^{1,2}$, H Honkanen$^{1,3}$ and C A Salgado$^{4,5}$}

\address{$^1$Department of Physics, P.O. Box 35, FI-40014 University of Jyv\"askyl\"a, Finland}
\address{$^2$Helsinki Institute of Physics, P.O. Box 64, FIN-00014 University of Helsinki, Finland}
\address{$^3$The Pennsylvania State University, 104 Davey Lab, University Park, PA 16802, USA}
\address{$^4$Departamento de F\'\i sica de Part\'\i culas and IGFAE, Universidade de Santiago de Compostela, Galicia-Spain}
\address{$^5$Physics Department, Theory Unit, CERN, CH-1211 Gen\`eve 23, Switzerland}

\ead{ilkka.helenius@jyu.fi}

\begin{abstract}
As all the globally fitted nuclear PDFs (nPDFs) have been so far impact parameter independent, it has not been possible to calculate the hard process cross sections in different centrality classes consistently with the global analyses. In \cite{Helenius:2012wd} we have offered a solution to this problem by determining two spatially dependent nPDF sets, \texttt{EPS09s} and \texttt{EKS98s}, using the $A$-systematics of the earlier global fits EPS09 and EKS98 and an assumption that the spatial dependence can be written as a power series of the nuclear thickness function. For a data comparison, we have calculated the nuclear modification factor of inclusive neutral pion production in d+Au collisions at RHIC in four centrality bins at midrapidity and compared these to a PHENIX measurement. In addition, we have also performed a similar calculation for inclusive photon production in d+Au collisions at RHIC.
\end{abstract}

\section{Introduction}

In the collinear factorization framework \cite{Collins:1989gx,Brock:1993sz} the cross section for inclusive $k$ production in a hard process of a heavy ion collision between nuclei $A$ and $B$ can be computed as
\begin{equation}
\mathrm{d} \sigma^{AB \rightarrow k + X} = \sum\limits_{i,j,X'} f_{i}^A(x,Q^2) \otimes f_{j}^B(x,Q^2) \otimes \mathrm{d}\hat{\sigma}^{ij\rightarrow k + X'} + {\cal O}(1/Q^2),
\label{eq:factorization}
\end{equation}
where $f_i^A$ ($f_j^B$) is the process-independent nuclear parton distribution function (nPDF) for a parton flavor $i$ ($j$) in the nucleus $A$ ($B$) and $\mathrm{d}\hat{\sigma}^{ij\rightarrow k + X'}$ represents the partonic pieces which can be computed using perturbative QCD (pQCD). Similarly as the free proton PDFs, also the nPDFs can be determined via a global analysis considering as many different processes as possible. 
Usually the nPDFs are determined in terms of the free proton PDFs, $f_{i}^{N},(x,Q^2)$ and the nuclear modification of the PDFs, $R_{i}^{A}(x,Q^2)$, as
\begin{equation}
f_{i}^{A}(x,Q^2) = R_{i}^{A}(x,Q^2) \, f_{i}^{N}(x,Q^2).
\end{equation}
So far all the globally analyzed nPDFs have been spatially independent. However, it is reasonable to assume that as the nucleus itself is not spatially uniform, also the nuclear modifications of the PDFs should somehow depend on the position of the nucleon inside the nucleus. In Ref.~\cite{Helenius:2012wd} we have addressed this issue by considering the $A$-systematics of two globally fitted nPDF sets, EPS09 \cite{Eskola:2009uj} and EKS98 \cite{Eskola:1998df} using the framework discussed in the next section.

\section{Framework}

\begin{figure}[htb]
\begin{minipage}[t]{0.49\linewidth}
\centering
\includegraphics[width=\textwidth]{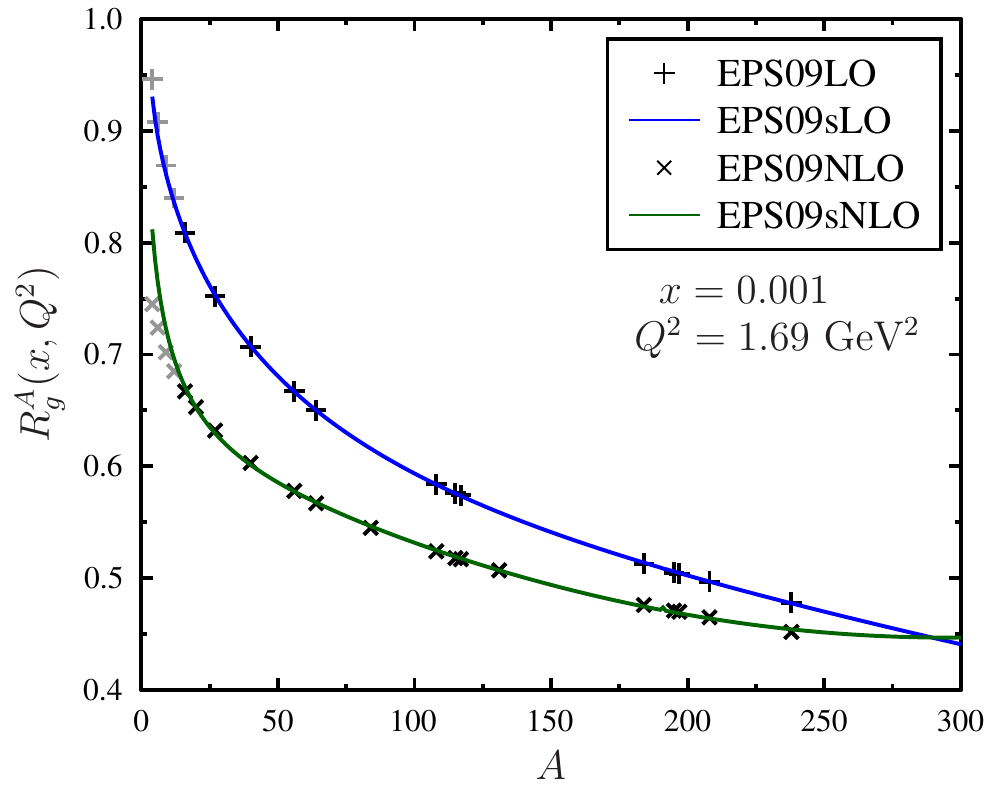}
\caption{The $A$-dependence of $R_g^A(x,Q^2)$ at fixed $x$ and $Q^2$ values from the central sets of EPS09NLO (crosses) and LO (pluses) and from EPS09sNLO (green) and LO (blue). From \cite{Helenius:2012wd}.}
\label{fig:R_g_A_fits}
\end{minipage}
\hspace{0.02\linewidth}
\begin{minipage}[t]{0.49\linewidth}
\centering
\includegraphics[trim = 10pt 0pt 0pt 0pt, clip, width=\textwidth]{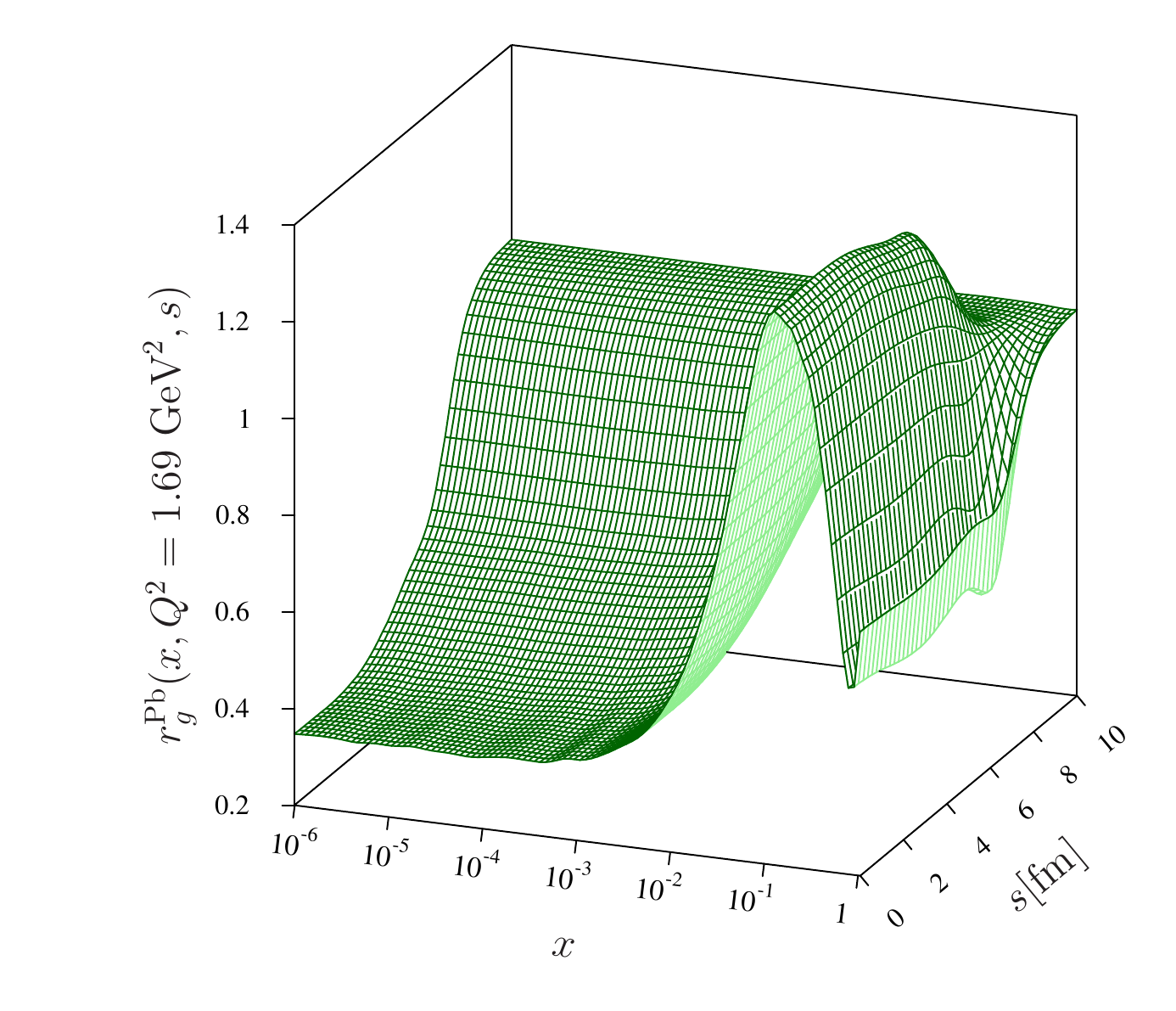}
\caption{The gluon modification in a lead nucleus $r_g^{\rm Pb}(x,Q^2,s)$ from EPS09sNLO as a function of $x$ and $s$ at the EPS09 initial scale. From \cite{Helenius:2012wd}.}
\label{fig:R_g_3d}
\end{minipage}
\vspace{-10pt}
\end{figure}

First we introduce a nuclear modification of the PDFs, $r_{i}^{A}(x,Q^2,\mathbf{s})$, which now depends also on the transverse position $\mathbf{s}$ of the nucleon inside the nucleus. For this we require that the spatial average of the quantity gives back the spatially independent modification,
\begin{equation}
R_{i}^{A}(x,Q^2) = \frac{1}{A}\int \mathrm{d}^2 \mathbf{s} \,T_A(\mathbf{s})\,r_{i}^{A}(x,Q^2,\mathbf{s}),
\end{equation}
for which we take the values from earlier global fits (EKS98 or EPS09). This, however, does not restrict the form of the spatial dependence in any way, so we have to make an assumption for that. Motivated by the shadowing region at small $x$, we assume that the $r_{i}^{A}(x,Q^2,\mathbf{s})$ can be written as a power series of the nuclear thickness function $T_A(\mathbf{s})$:
\begin{equation}
r^{A}_{i}(x,Q^2,\mathbf{s}) = 1 + \sum_{j=1}^{n} c_j^i(x,Q^2)\left[ T_A(\mathbf{s})\right]^{j},
\label{eq:series}
\end{equation}
where now the fit parameters $c_j^i(x,Q^2)$ for each parton flavor depend only on $x$ and the scale $Q^2$ but not on $A$. This is important for correct mapping of the spatial dependence with the $A$-dependence of the globally fitted $R_i^A$. In practice we obtain the values for our fit parameters $c_j^i(x,Q^2)$ by minimizing the $\chi^2$ defined as
\vspace{-5pt}
\begin{equation}
\chi^2_i(x,Q^2) = \sum_A \left[\frac{R^{A}_{i}(x,Q^2) - \frac{1}{A}\int \mathrm{d}^2 \mathbf{s}\, T_A(\mathbf{s})r^{A}_{i}(x,Q^2,\mathbf{s})}{W^{A}_{i}(x,Q^2)} \right]^2
\end{equation}
in a $(x,Q^2)$ grid for all the parton flavors. The weight factor $W^{A}_{i}(x,Q^2)$ is set to 1 ($1-R^A_i(x,Q^2)$) for the EPS09 (EKS98) analysis. For the EPS09 analysis we perform this fitting also for the 30 error sets both in leading (LO) and next-to-leading order (NLO). As can be seen from figure~\ref{fig:R_g_A_fits}, the $A$-systematics of $R_i^A$ is very well reproduced with the power series ansatz when we consider $A \ge 16$. In our analysis we found out that the first four non-trivial terms of the power series in equation~\ref{eq:series} are enough for an accurate fitting in the whole $x$ and $Q^2$ region, for all the different sets considered.

After the values for the fit parameters $c_j^i(x,Q^2)$ are obtained through this fitting procedure, we can calculate the $r_{i}^{A}(x,Q^2,\mathbf{s})$ and determine the new spatially dependent nuclear modification sets, \texttt{EPS09s} and \texttt{EKS98s}, in which ''\texttt{s}'' stands for ''spatial''. These are now published in our website\footnote{\url{https://www.jyu.fi/fysiikka/en/research/highenergy/urhic/nPDFs}}. To illustrate the spatial dependence of the new nPDFs, the gluon modification from EPS09sNLO for lead nucleus is plotted in figure~\ref{fig:R_g_3d} as a function of $x$ and the transverse distance $s=|\mathbf{s}|$. The general feature is that the nuclear effects are larger in the dense center (small $s$) of the nucleus and smaller at the sparse edge (large $s$).

\section{Applications}

The nuclear effects of an observable can be quantified using the nuclear modification factor $R_{AA}$. The improvement here is that now with the new nPDF sets one can compute the $R_{AA}$ also for different centrality classes in a manner which is consistent with the globally fitted nuclear modifications of the PDFs. Detailed instructions of the implementation of these new nPDFs can be found in Ref.~\cite{Helenius:2012wd}.

\begin{figure}[htb]
\centering
\includegraphics[width=0.9\textwidth]{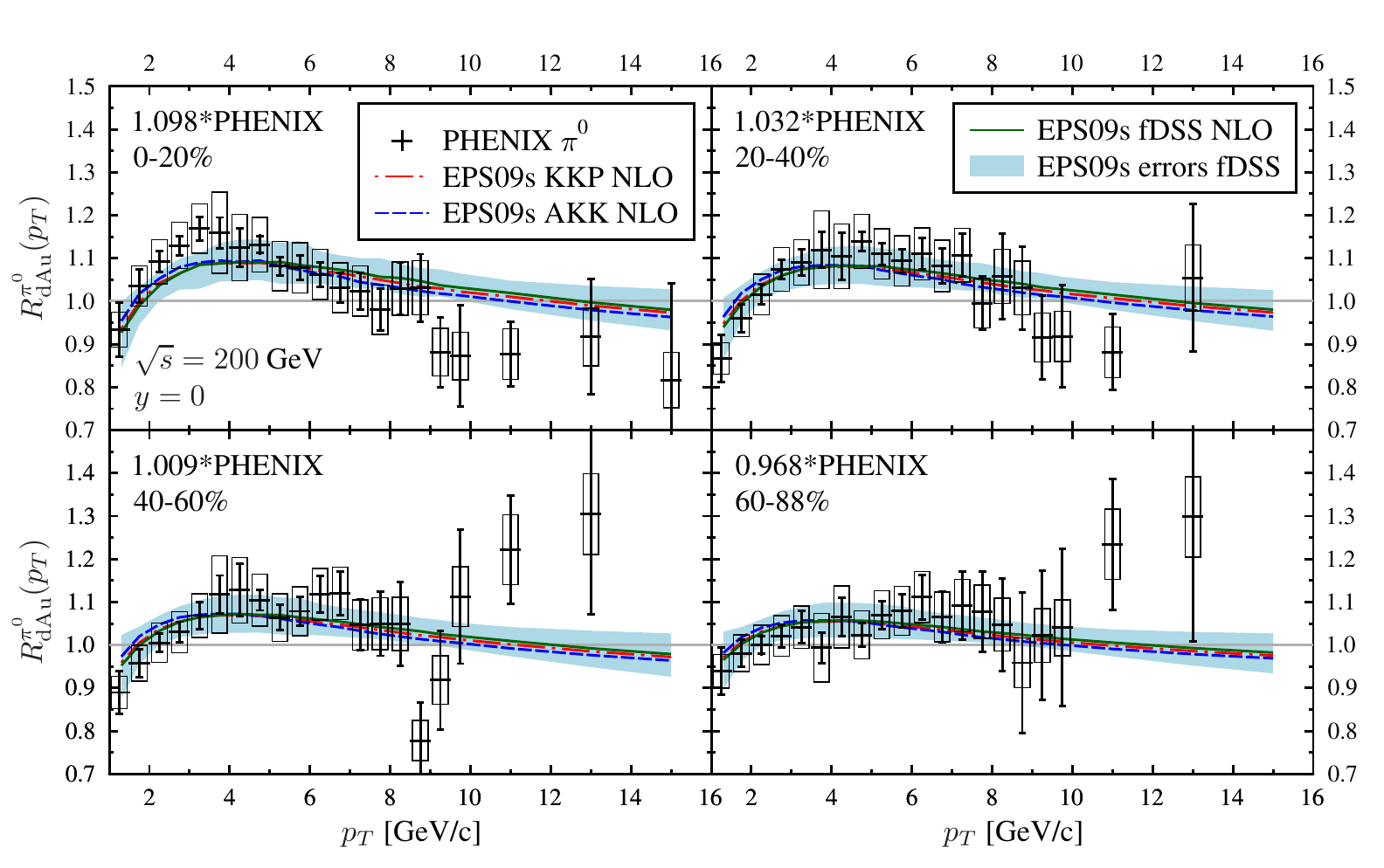}
\vspace{-14pt}
\caption{The nuclear modification factor for inclusive $\pi^0$ production in d+Au collisions at RHIC in four centrality classes at midrapidity, calculated with the EPS09s nPDFs and different FFs. The data are from PHENIX \cite{Adler:2006wg}. From \cite{Helenius:2012wd}.}
\label{fig:R_dAu_pi0}
\end{figure}

In figure \ref{fig:R_dAu_pi0} we have plotted the nuclear modification factor $R_{\rm dAu}^{\pi^0}$ for inclusive $\pi^0$ production in d+Au collisions at $\sqrt{s_{NN}} = 200\,\rm{GeV}$ and $y = 0$ as a function of $p_T$ in four centrality classes. The calculations are done in NLO (with the \texttt{INCNLO}-package\footnote{\url{http://lapth.in2p3.fr/PHOX_FAMILY/readme_inc.html}}) using the CTEQ6M PDFs \cite{Pumplin:2002vw} with the EPS09s modifications \cite{Helenius:2012wd} and three different fragmentation functions (FFs), KKP \cite{Kniehl:2000fe}, AKK \cite{Albino:2008fy}, and fDSS \cite{deFlorian:2007aj}. The uncertainty bands are calculated from the error sets of EPS09s with the fDSS FFs. The centrality classes are defined in terms of impact parameter intervals, which are calculated using the optical Glauber model. The PHENIX datapoints \cite{Adler:2006wg} have been scaled by overall factors which all are consistent with the overall normalization uncertainties given by the experiment. The corresponding results for the forthcoming p+Pb collisions at the LHC can be found in \cite{Helenius:2012wd}.
\vspace{-29pt}

\begin{figure}[tbh]
\centering
\includegraphics[width=0.9\textwidth]{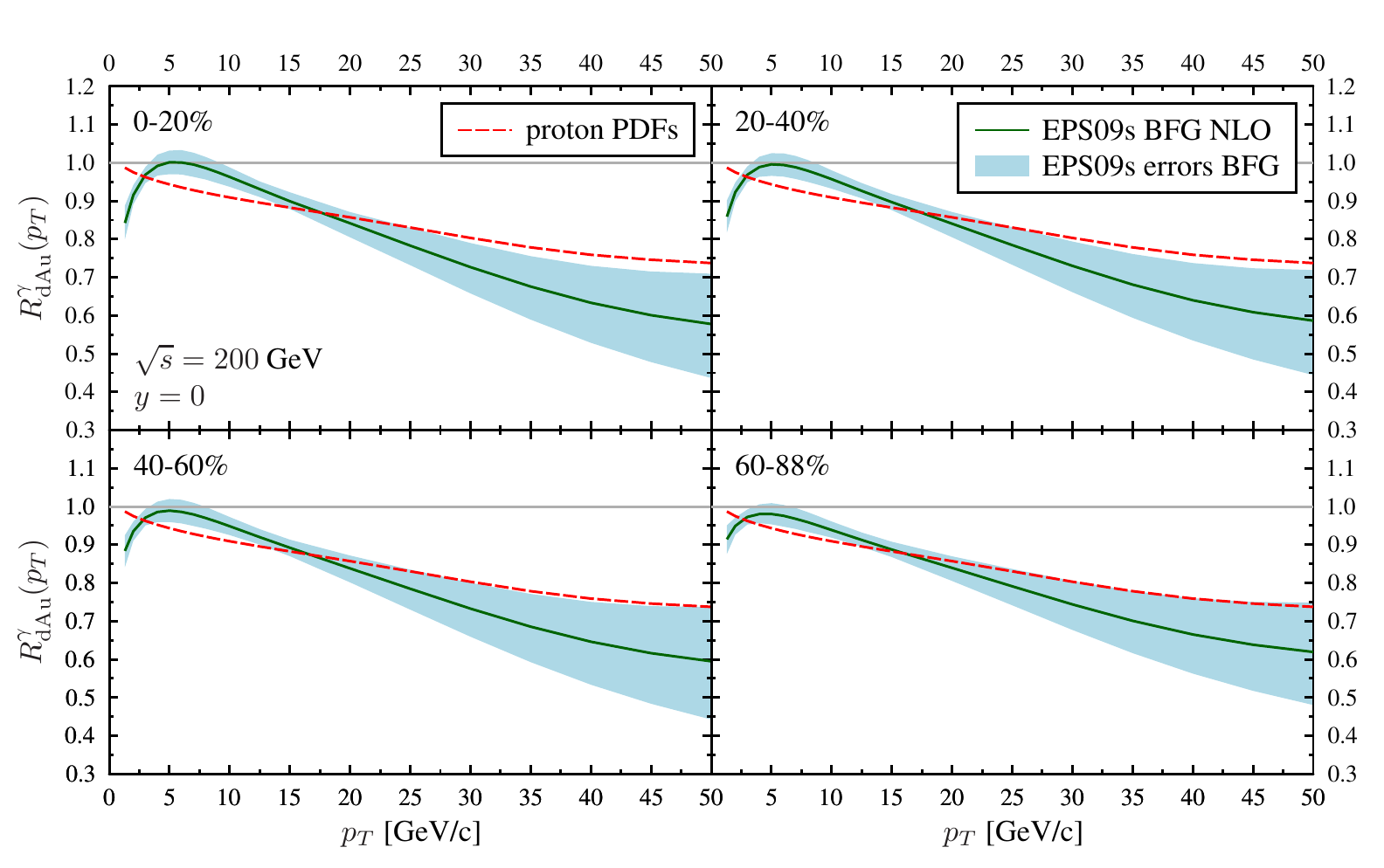}
\vspace{-14pt}
\caption{The nuclear modification factor for inclusive $\gamma$ production in d+Au collisions at RHIC in four centrality classes at midrapidity, computed with nPDFs (EPS09s) and proton PDFs (CTEQ6M). From \cite{HeleniusEskola}.}
\label{fig:R_dAu_gamma}
\end{figure}

We have also performed similar calculations for inclusive photon production, $R_{\rm dAu}^{\gamma}$, in d+Au collisions at RHIC, which are shown in figure \ref{fig:R_dAu_gamma}. The results for the minimum bias nuclear modification factor for this process, calculated with different nPDFs, can be found in Ref.~\cite{Arleo:2011gc}. The inclusive photons consists of two components: direct and fragmentation. For the fragmentation component we have used the BFG (set II) FFs \cite{Bourhis:1997yu}, otherwise the setup is the same as for the $\pi^0$'s above. To quantify the isospin effect, we have also plotted the $R_{\rm dAu}^{\gamma}$ using the free nucleon PDFs in each panel. Similarly as for pions, also for the photons we can see that the nuclear modifications are larger for central collisions than for peripheral collisions. 
\vspace{-22pt}
\vspace{-15pt}

\ack

\small{
I.H. and K.J.E. thank the Magnus Ehrnrooth Foundation and Academy of Finland (Project 133005) for financial support. 
C.A.S. is supported by the European Research Council grant HotLHC ERC- 2001-StG-279579, by Ministerio de Ciencia e Innovaci\'on of Spain, and by Xunta de Galicia.
H.H. is supported by the U.S. Department of Energy under Grant  DE- FG02-93ER40771.
}


\section*{References}
\bibliographystyle{iopart-num}
\bibliography{Vietnam2012_proceedings}

\end{document}